\documentclass[12pt]{article}
\usepackage{epsfig}
\newcommand{\R}{I\!\!R}

\newcommand{\noi}{\noindent}

\newcommand{\la}{\langle}
\newcommand{\ra}{\rangle}

\newcommand{\rhotilde}{\tilde{\rho}}
\newcommand{\rhoeq}{\rho_{eq}}
\newcommand{\lambdadot}{\dot{\lambda}}
\newcommand{\lambdahat}{\hat{\lambda}}
\newcommand{\ahat}{\hat{a}}
\newcommand{\Mhat}{\hat{M}}
\newcommand{\Ghat}{\hat{G}}
\newcommand{\Khat}{\hat{K}}
\newcommand{\Psihat}{\hat{\Psi}}

\renewcommand{\d}{\partial}
\newcommand{\half}{\frac{1}{2}}

\renewcommand{\L}{{\cal L}}
\renewcommand{\H}{{\cal H}}

\begin{document}

\title{{\bf Best-fit quasi-equilibrium ensembles: a general
approach to statistical closure of underresolved Hamiltonian
dynamics } }
\author{ Bruce Turkington
\\
 University of
Massachusetts Amherst\\  and \\ 
Petr Plech\'a\v{c} \\ University
of Tennessee Knoxville and \\ Oak Ridge National Laboratory}  
\date{}
\maketitle

\begin{abstract}
A new method of deriving reduced models of Hamiltonian dynamical systems is
developed using techniques from optimization and statistical estimation.   
Given a set of resolved variables that define a model reduction,   
the quasi-equilibrium ensembles associated with the resolved variables
are employed as a family of trial probability densities on phase space.   
The residual that results from submitting these trial densities to the Liouville equation
is quantified by an ensemble-averaged cost function
related to the information loss rate of the reduction.    
From an initial nonequilibrium state,  the statistical state of the system at any later time 
is estimated by minimizing the time integral of the cost function over paths of trial densities.  
Statistical closure of the underresolved dynamics is obtained at the level of the value function,
which equals the optimal cost of reduction with respect to the resolved variables, and     
the evolution of the estimated statistical state is deduced from the Hamilton-Jacobi equation
satisfied by the value function.     
In the near-equilibrium regime, or under a local quadratic approximation in the
far-from-equilibrium regime,  this best-fit  closure is governed by a differential
equation for the estimated state vector coupled to a  Riccati differential
equation for the Hessian matrix of the  value function.    
Since memory effects are not explicitly included in the trial densities,  
a single adjustable parameter is introduced into the cost function
to capture a time-scale ratio between resolved and unresolved motions.   
Apart from this parameter,  the closed equations for the resolved variables
are completely determined by the underlying deterministic dynamics.   

\end{abstract}

\noi {\it Key Words and Phrases: nonequilibrium statistical mechanics,
turbulence closure, model reduction, statistical estimation, optimization, 
Hamilton-Jacobi equation}

\section{Introduction}

Complex nonlinear dynamical systems with many interacting degrees of freedom,
or many coupled modes of motion, are formulated throughout the sciences 
for the purpose of making reliable predictions about the evolution of system states.   
But the practical usefulness of these high-dimensional models is limited 
unless they are combined with some kind of model reduction.   
Indeed, what is usually desired from a model of a complex system is
a quantitative description of some robust, collective behavior.   Such
a description is not necessarily furnished by individual solution trajectories,  
owing to the generic presence of deterministic chaos.      
Moreover, in realistic problems 
neither the specification of initial states nor the measurement of evolved states is
exact.  It is generally desirable, therefore, to select some subset of the system's dynamical
variables and declare them to be resolved, or relevant, variables, and to seek 
 a reduced dynamical description in terms of them.    For instance,
in a spatially-extended system the resolved variables could furnish
a coarse-grained description of the fully resolved, fine-grained dynamics.
Aside from the practical considerations that constrain resolution in numerical simulations, 
the selection
of these resolved variables is normally determined by two considerations: (1) 
how initial states are prepared and evolved states are observed,
and (2)  whether it is possible to achieve an approximate closure of the 
dynamics in terms of the resolved variables.      
In most instances there is no perfect selection.  Also, there is a competition between
these two criteria, because the first  is more easily satisfied by a few resolved variables,
while the second  is better achieved by many.  
In this light,  the modeler is often confronted with 
the general problem of deriving a reduced system of governing equations 
for a selected, though not unique, set of resolved variables.

This model reduction procedure is necessarily statistical, since
it relegates all unresolved variables to a probabilistic description.  
A systematic approach to model reduction therefore naturally makes 
recourse to the methods of statistical mechanics, which furnishes 
a collection of methodologies for
deriving macroscopic behavior from microscopic dynamics
 \cite{Balescu,Balian,Chandler,Katz}.     But this
field has traditionally developed in the narrower context of deriving
thermodynamical properties of matter from the attributes and
interactions of a huge number of elementary constituents.  
Moreover, since reduced equations governing statistically-averaged resolved
variables for a complex system are analogous to transport equations
for thermodynamics variables, the most pertinent  methods are those 
of nonequilibrium statistical
mechanics \cite{Balescu,Keizer,Zwanzig}.  Unlike equilibrium
statistical mechanics, which is a general theory resting on the secure
foundation of Gibbs ensembles, nonequilibrium statistical mechanics is
still an emerging field whose various branches treat particular
physical processes and whose methods require special mathematical
simplifications.  As a result, there remain many problems in model
reduction on which little progress has been made because they pertain
to phenomena lying outside the range of existing nonequilibrium
theory.  For instance, many aspects of turbulence modeling suffer from
the lack of systematic approximations  that are analytically justified and 
computationally tractable,
and it is mainly for this reason that the design of reliable
statistical closures for turbulent dynamics remains such a challenging open
problem.

In this paper we propose a new approach to model reduction for
complex deterministic dynamics.   In order to maintain contact with
the fundamental notions of statistical mechanics, we focus on 
classical Hamiltonian systems.  The essence of our approach
to statistical closure is an optimization procedure in which we
seek  computationally tractable approximations to solutions
of the Liouville equation, which governs the exact evolution of 
probability density on phase space \cite{Balescu,Katz,Zwanzig}.   
Instead of manipulating exact but computationally inaccessible solutions
of the Liouville equation, we use trial
probability densities on phase space that form a parametric statistical model 
for which the given resolved variables are minimal sufficient statistics \cite{CB}.
With respect to this family of trial densities, we
seek paths in the statistical parameter space that have minimal lack-of-fit  to 
the Liouville equation.   Specifically, we minimize the 
time integral of an ensemble-averaged, quadratic cost function of the 
Liouville residual over those paths of trial densities
which connect an initial nonequilibrium state to an estimated state at any later time.    
In this way we obtain a closure that is
optimal relative to the resolved variables selected for  model reduction. 

To complete the derivation of the statistical closure,  
we deduce the set of differential equations governing the evolution
of the estimated parameters for the statistical model.  
To do this we introduce  
the value function for our best-fit minimization problem, which is a function
on the statistical parameter space that gives the optimal 
cost of reduction.   As is known in optimization theory \cite{FR,GF}, this value function
satisfies a Hamilton-Jacobi equation with a Hamiltonian that is conjugate to
the Lagrangian cost function.   The desired equations governing the 
best-fit statistical state are therefore determined by the solution propagated by that Hamilton-Jacobi
equation.  Finally, computationally tractable simplifications of them are systematically 
extracted under some further approximations.        

We restrict our attention here to the relaxation problem for 
Hamiltonian dynamical systems.    Apart from mild regularity and growth
conditions, we put no restrictions on the Hamiltonian defining the
underlying complex dynamics or on the set of resolved variables defining
the model reduction.  
In this  general context  we derive reduced equations that approximate the
evolution toward equilibrium of ensemble-averaged resolved variables from 
incompletely specified, nonequilibrium initial values.  
Our purpose is to demonstrate how an irreversible
statistical closure can be inferred  from an underlying
dynamics that is deterministic, reversible and conservative.    
Our approach to  nonequilibrium statistical behavior differs from much of the current
literature in that we do not assume that the 
microscopic dynamics is a known stochastic processes, nor do we interpose a
stochastic model between the closed reduced dynamics and the
Hamiltonian dynamics.    While our 
general method could be adapted to forced and dissipative systems, 
we do not consider these systems in the present paper, nor do we address
the much studied question of stationary statistical states for those dynamical
systems.    

For the sake of definiteness, we limit our presentation to the
particular case of the best-fit estimation strategy in which the trial
probability densities are quasi-equilibrium, or quasi-canonical,
ensembles \cite{Zubarev,LVR,Penrose}.  Namely, we use densities 
\begin{equation}   \label{qe-density}
\rhotilde(z; \lambda) = \, 
           \exp [ \lambda^* A(z) - \phi(\lambda) ] \, \rho_{eq}(z) \, ,
\end{equation} 
in which
\[
\phi(\lambda) \, =\, 
    \log \int_{\Gamma_n} \exp ( \lambda^* A(z) ) \,  \rho_{eq}(z) dz \; .
\] 
The $m$-vector $A=(A_1, \ldots , A_m)$  consists of the resolved variables
in the reduced model, and  $\rho_{eq}$ denotes an  equilibrium
probability density.      The family of densities (\ref{qe-density})
is parameterized by the real $m$-vector $\lambda = (\lambda^1, \ldots ,
\lambda^m) \in \R^m$.   
Here and throughout the paper, $*$ denotes the transpose of a real
vector or matrix (so that in these formulas, $\lambda^*A = \lambda^1
A_1 + \cdots + \lambda^m A_m$); $z$ denotes the generic point in the
phase space $\Gamma_n \subset \R^{2n}$, and $dz$ is the element of phase volume on
$\Gamma_n$.  For the purposes of our general development, the resolved variables, $A_k $, 
may be any independent,  real-valued, smooth functions on  $\Gamma_n$ 
that are not conserved by the Hamiltonian dynamics, and     
 $\rho_{eq}$, may be any invariant density, such as
the canonical Gibbs ensemble.    From the perspective of statistical inference, 
$\rhotilde (z;\lambda)$ is an exponential
family  on $\Gamma_n$ with the natural (or canonical) parameter $\lambda$, 
and the random vector $A$ is a minimal sufficient statistic for this family \cite{CB}.

In the reduced model the ensemble mean of $A$ with respect to
$\rhotilde$ constitutes the macrostate 
\[
a = (a_1, \ldots, a_m) \, = \, \la A \, | \rhotilde \, \ra \, = \, 
     \int_{\Gamma_n} A (z) \, \rhotilde(z;\lambda) \, dz \, .    
\]
The parameter vector $\lambda$ is dual to the macrostate vector $a$, and
the convex function $\phi$ determines a one-to-one correspondence
between $\lambda$ and $a$ through $a=\partial \phi / \partial
\lambda$.    The choice of the statistical model (\ref{qe-density})
defines a configuration space, $\R^m$, for the reduced model, the generic
point of which is $\lambda$.   
The desired reduced dynamics is therefore characterized
either by the evolution of $\lambda$ or equivalently by the evolution
of $a$.  This point of view is familiar from the information-theoretic
approach to nonequilibrium statistical mechanics
\cite{Jaynes1,Jaynes2,Katz, Zubarev,Dougherty,LVR}.

Our goal is to estimate the macrostate vector $a(t_1)$ at any time
$t_1$ after an initial time $t_0$ at which a nonequilibrium initial state 
is specified.   For simplicity in this Introduction, we assume that
the initial density, $\rho(z,t_0) = \rhotilde(z;\lambda_0)$, is a quasi-equilibrium
ensemble of the form (\ref{qe-density}) 
corresponding to a specified $\lambda(t_0) =
\lambda_0$, with $a(t_0) = a_0$;  in the body of the paper we give a
formulation in which the initial statistical state is incompletely specified.     
The exact density $\rho(z,t)$ at later times $t>t_0$ solves the
Liouville equation, which can be written in exponential representation as
\[
\frac{\partial \log \rho}{\partial t} \,+\, L \log \rho \;= \, 0   \, , 
  \]
where $L = \{ \cdot , H \}$ denotes the Liouville operator
associated to the Poisson bracket $\{ \cdot , \cdot \}$.  As 
$\rho$ evolves it loses its quasi-equilibrium form and
develops a more intricate structure than the trial probabilities
$\rhotilde$.  The key to an effective statistical closure with respect
to the given $m$-vector $A$ of resolved variables is therefore to devise a good
approximation to $\rho$ within the quasi-equilibrium family
$\rhotilde$.  To this end, we calculate the residual with respect to the Liouville equation 
of a  trial density $\rhotilde(z;\lambda(t))$ along any smooth path $\lambda(t)$
in the statistical parameter space $\R^m$:
\[
R \, = \, \frac{\partial \log \rhotilde}{\partial t} \,+\, L \log \rhotilde  \;= \;
    \lambdadot^* (A-a) \, + \, \lambda^* L A \, .
\]
This Liouville residual $R = R(z;\lambda,\lambdadot)$, or lack-of-fit
of the trial density $\rhotilde(z;\lambda)$ to the Liouville equation,  has an
interpretation as the instantaneous rate of information loss within the statistical model.
We therefore base our estimation strategy
for $a(t_1)$ on minimizing $R$ in an appropriate norm over the time horizon of estimation,
$t_0 \le t \le t_1$.   Namely, 
we evaluate the following time-integrated, ensemble-averaged  cost functional:
\begin{equation}   \label{cost-intro}
\int_{t_0}^{t_1} \L(\lambda, \lambdadot) \, dt  \,=\,  \frac{1}{2} \int_{t_0}^{t_1} 
         \la ( P_{\lambda} R )^2 \, | \rhotilde(\lambda)  \ra \, + 
              \, \epsilon^2 \la ( Q_{\lambda} R)^2 \, | \rhotilde(\lambda)   \ra \; dt
\end{equation}   
where $P_{\lambda}$ is the orthogonal projection of $L^2(\Gamma_n, \rhotilde(\lambda))$
onto the span of the resolved vector $A$, and $Q_{\lambda} = I  - P_{\lambda}$
is the complementary projection.    The constant $\epsilon \in (0,1]$ is an adjustable
parameter in our closure scheme, which assigns relative weights to the resolved 
and unresolved components of the  Liouville residual.       

The integrand in the cost functional (\ref{cost-intro}) may be viewed 
as a Lagrangian $\L(\lambda,\lambdadot)$ on the configuration space $\R^m$.
In that light our best-fit strategy for closure is determined by an analogue to
the classical principle of least action.    
We therefore use Hamilton-Jacobi theory \cite{Arnold,GF,Lanczos,Evans}
to deduce the governing equations for our closed reduced dynamics.   That is,
we introduce the value function (or principal function) 
\begin{equation}    \label{value-a}
v(\lambda_1,t_1) \; = \; \min_{\lambda(t_1)=\lambda_1} \int_{t_0}^{t_1} \L(\lambda, \lambdadot) \, dt 
\;\;\;\;\;  \mbox{ subject to } \;\; \lambda(t_0) = \lambda_0 \, , 
\end{equation}  
which associates an optimal lack-of-fit to an arbitrary terminal state $\lambda_1$ at a terminal
time $t_1$.  This minimization is over admissible paths  $\lambda(t), \; t_0 \le t \le t_1$, that
connect the specified initial state $\lambda_0$ to  $\lambda_1$.     
The value function, $v=v(\lambda,t)$, satisfies the associated Hamilton-Jacobi equation
\begin{equation}  \label{hj-intro}
\frac{\d v}{\d t} + \H(\lambda, \frac{\d v}{\d \lambda}) \; = 0   
\end{equation}  
in which $\H(\lambda, \mu)$ is the Legendre transformation of $\L(\lambda,\lambdadot)$.   
The conjugate canonical  variable, $\mu$,  has an interpretation as
the irreversible part of the flux of the
macrostate $a = \la A | \rhotilde \ra$, in the sense that
\[
\mu = \frac{\d \L}{\d \lambdadot} \, = \, \la R A \, | \, \rhotilde \ra \,=\, 
    \frac{d}{dt} \la A | \rhotilde \ra - \la L A | \rhotilde \ra \; . 
 \]
 The Hamilton-Jacobi equation propagates the value function forward in time
 from the singular initial condition that all paths emanate from $\lambda_0$.   
 We designate the best-fit estimate of the statistical state at any time $t$  
 to be the minimizer $\lambdahat(t)$ of $v(\lambda,t)$;  
the corresponding best-fit macrostate is therefore 
$\ahat(t) = \la A | \rhotilde(\lambdahat(t)) \ra =  \d \phi / \d \lambda (\lambdahat(t))$.

The  reduced dynamics governing the evolution of the best-fit macrostate can be deduced
from the minimizing property of  $\lambdahat(t)$ and the Hamilton-Jacobi equation; 
the main result is
\begin{equation}   \label{reduced-eqn-intro}
\frac{d \ahat}{dt} \;=\; \la L A \, | \, \rhotilde (\lambdahat) \ra \,- \, 
    \epsilon^2 \, D^2 \phi(\lambdahat)  
         \left[ D^2 v (\lambdahat,t) \right]^{-1} 
          \frac{\d w}{\d \lambda} (\lambdahat) \, ,  
\end{equation}
where $w(\lambda) = \la \, [Q_{\lambda} L (\lambda^*A )]^2 \, |\, \rhotilde (\lambda) \, \ra$,
and $D^2 \phi$ and $D^2 v$ are Hessian matrices with respect to $\lambda$.  
 The right hand side of this equation separates into an reversible term, which has a standard form,
 and an irreversible term scaled by $\epsilon^2$, which has a novel form.   The 
 irreversible part of the flux in (\ref{reduced-eqn-intro}) has a generalized
 gradient structure with a potential-type
 function $\epsilon^2 w(\lambda)$ that quantifies the influence of unresolved fluctuations.
 Besides being scaled by $\epsilon^2$ the irreversible term also depends on $\epsilon$ implicitly through the value function $v=v(\lambda,t;\epsilon)$.   

The defining minimization over paths on the entire
time interval of estimation partially compensates for the fact that
the trial densities are memoryless, quasi-equilibrium ensembles.    
This feature of the best-fit closure is its primary advantage as a computational method,
since no trajectories of the underlying Hamiltonian need be computed. 
However, this feature also requires that the scale factor
$\epsilon$ be introduced into the cost function and  be adjusted to give 
the correct dissipation rate in the reduced dynamics.    It is not surprising that such adjustment
is needed in light of known projection methods \cite{Zwanzig,Prigogine_etal,CHK1,CHK2}, 
which furnish various expressions for the dissipative term in (\ref{reduced-eqn-intro}).
These expressions involve time convolutions with respect to a memory kernel for
autocorrelations with respect to a projected Liouville propagator, 
essentially $e^{t Q L}$ in our notation.  
The cost functional (\ref{cost-intro}) incorporates  a minimal representation of the memory of
unresolved fluctuations by means of a single real parameter $\epsilon$.
This approach dispenses with the computation of any memory kernel
at the expense of an adjustable parameter that must be tuned empirically.   

A further approximation of the reduced dynamics (\ref{reduced-eqn-intro}) is desirable 
because evaluation of  the value function requires solving the Hamilton-Jacobi
equation, which is computational burdensome except when $m$ is small.    Accordingly,
we derive a more explicit closure scheme for the pair $(\lambdahat(t), \Mhat(t))$, where
$\Mhat(t) = D^2v(\lambdahat(t),t)$ denotes the Hessian matrix of the value function
at the best-fit state.     
To do so we make the  local quadratic approximation
$v(\lambda,t) \approx v(\lambdahat(t),t) + 
(1/2) [ \lambda -\lambdahat(t) ]^* \Mhat(t)  [ \lambda -\lambdahat(t) ]  \;$ 
in a neighborhood of $\lambdahat$, and we find that the Hamilton-Jacobi equation
reduces to a coupled system for $\lambdahat$ and $\Mhat$, in which 
$\Mhat$ satisfies a matrix Riccati differential equation with coefficient matrices 
that depend on $\lambdahat$.  
The local quadratic approximation used to derive these closed reduced equations 
from the general best-fit theory has the character of a truncation of
a closure hierarchy at the second order.   In this sense it is a rational method
of acquiring a computationally tractable  statistical 
closure that could be applied far from equilibrium.   

We also show that, in the near equilibrium regime
under the standard linear response approximations \cite{Chandler,Zwanzig}, this local
quadratic representation of the value function is valid globally and the
matrix Riccati equation has constant coefficients.   Thus, for near equilibrium
behavior we obtain a simpler system of closed reduced equations, which is comparable 
to the phenomenological equations of linear irreversible thermodynamics 
\cite{dGM,Ottinger},
but with a time-dependent matrix of transport coefficients that is determined 
by the solution to the Riccati equation.     

Our approach seems to have no antecedent in the literature.
While we build on the ideas of Jaynes \cite{Jaynes1,Jaynes2} and
Zubarev \cite{Zubarev}, as well as later workers
\cite{Robertson,LVR,ZMR}, in that we utilize quasi-equilibrium
densities, which maximize entropy subject to instantaneous macroscopic
constraints,  our strategy of minimizing a time-integrated cost function
for the Liouville residual over paths of these densities appears to be new.   
Furthermore, our closed reduced equations have a different format from
that of other theories.    In particular, the differential equation for the estimated statistical
state $\lambdahat(t)$ is coupled to a differential equation for the Hessian matrix,
$\Mhat(t)$,  which quantifies the information content in the estimate.   
In this sense our theory simultaneously estimates the evolving macrostate
and quantifies the uncertainty inherent in that estimate.    The
format of our closed reduced equations, therefore, resembles that of continuous-time
optimal filtering theory \cite{BH,Davis,McGarty,MHG}, 
and in the near equilibrium regime there is a close connection
between our reduced equations and the Kalman-Bucy filter.    
Unlike the standard filtering problem, however, our statistical closure
updates the reduced model by accessing the unresolved components of the
deterministic equations of motion themselves, rather than assimilating some measurements.   
 
The outline of the paper is as follows.   We set the background for
our approach in Section 2, and then we present the best-fit closure
scheme in Sections 3 and 4.   This theoretical development  applies to
nonequilibrium states that may be far from equilibrium.  
In Sections 5 and 6 we specialize our general formulation to the
near-equilibrium version of our theory.    In the case when
all the resolved variables are symmetric under time reversal, we
show that our near-equilibrium
closure takes a form similar to that of linear irreversible thermodynamics.   
In Section 7 we give a heuristic analysis that supports the physical
interpretation of the adjustable parameter $\epsilon$ as a time-scale ratio.    

In a companion paper  \cite{PT} we address the computational
implementation of the general methodology 
developed in this paper, and we present comparisons of its
predictions against fully-resolved numerical simulations for a
particular Hamiltonian system.

\section{Background}

We consider a general Hamiltonian dynamics in canonical form
\begin{equation}  \label{hamilton_eqn}
\frac{dz}{dt} \,=\, J \nabla_z H(z) \;\;\;\; \mbox{ with } \;\;\;\; J
\,=\, \left( \begin{array}{ccc} O & I \\ -I & O
  \end{array}  \right) \, ,  
\end{equation}
where $z=(q,p)$ denotes a generic point in the phase space $\Gamma_n =
\R^{2n}$, where $n$ is the number of degrees of
freedom of the system.  We place no special restrictions on the
Hamiltonian $H$ other than it be a smooth function on $\Gamma_n$ with
a natural growth condition at infinity:  $ H(z) \ge b |z|^2 - c$, with $b>0, c
\ge0$, for large $|z|$.  Most of what follows holds for noncanonical Hamiltonian systems,
but we will restrict our development in this paper to classical
canonical systems for the sake of clarity.

We denote the phase flow for the Hamilton equations
(\ref{hamilton_eqn}) by $\Phi(t) : \Gamma_n \rightarrow \Gamma_n$, so
that
\[
z(t_1) \,=\, \Phi(t_1-t_0) (z(t_0))   \;\;\;\; \mbox{ for all }
  t_1 \ge  t_0 \, ,
\]
where $z(t)$ is any solution of (\ref{hamilton_eqn}).  This
deterministic phase flow $\Phi(t)$ is a volume-preserving
diffeomorphism of $\Gamma_n$ for all $t $, by Liouville's theorem.  
The invariant $2n$-volume on $\Gamma_n$ is denoted by $dz$.

We are interested in estimating or approximating the macroscopic
behavior of a few dynamical variables rather than
following the details of the microscopic dynamics (\ref{hamilton_eqn})
itself.  We therefore suppose that some resolved, or relevant, dynamical variables
are selected, and we seek a statistical closure in terms of these variables.
We assume that each dynamical variable $A_k$ is a
smooth real-valued function on $\Gamma_n$, and that the set $A_1,
\ldots , A_m$ is linearly independent.   We assemble them into the 
resolved vector $A=(A_1, \ldots ,A_m)$.   In principle there is no
restriction on $m$, but in practice the number $m$ of relevant
variables should be small compared to the dimension $n$ of the phase
space.    

The quality of any choice of resolved variables
is determined by the ability of the resulting reduced model to approximate 
the collective behavior of the system.   The assessment of any particular selection 
of resolved variables therefore first requires the formulation of a closure scheme.
Nonetheless, in most practical problems  there will be some natural
choices of $A$ in terms of which it is reasonable to expect a good
approximation.     We mention just two such physical systems: 
molecular dynamics and dispersive wave turbulence.  
For a coupled system of particles in which a few tagged particles
interact with many particles constituting a ``heat bath,"  the canonical variables of the 
tagged particles furnish natural resolved variables for reduction \cite{FKM,GKS}.  
For a nonlinear wave system with many interacting modes, a kinetic description
of the power spectrum (perhaps in some selected bands) is a customary reduction 
 for ``weak turbulence" theory \cite{ZLF,CM}.      
Even in these cases, though, improved approximations may result 
from expanding or otherwise modifying the set of resolved variables.   In light
of considerations of this kind, we proceed without putting any special
restrictions on the resolved vector $A$.

The evolution of any dynamical variable, $F$, resolved or not,
is determined by the equation
\begin{equation}  \label{poisson_eqn} 
\frac{dF}{dt} \,=\, \{ F , H \} \, ,
\end{equation}
where $\{F,H\} = (\nabla F)^* J \nabla H$ is the Poisson bracket
associated with the canonical Hamiltonian structure.  Indeed, the
statement that (\ref{poisson_eqn}) holds for all smooth functions $F$
on $\Gamma_n$ is equivalent to the Hamiltonian dynamics
(\ref{hamilton_eqn}).  Fundamentally, the problem of closure in terms
of the resolved variables $A_1, \ldots, A_m$ arises from the fact
that, except under very special circumstances, the derived
variables $\dot{A}_1 = \{A_1,H\}, \ldots, \dot{A}_m = \{A_m,H\}$ are
not expressible as functions of $A_1, \ldots, A_m$.   For instance, in
the exceptional case when 
$\dot{A} = \Omega A$ identically on $\Gamma_n$ for
some $m \times m$ constant matrix $\Omega$, a
deterministic closure is immediate from (\ref{poisson_eqn}).
We are however interested in the generic case, and hence we  adopt a statistical description
defined by that evolving
probability measure $p(dz,t)$ on the phase space $\Gamma_n$ which is
induced by the phase flow; namely, 
\[
\int_{\Phi(t_1-t_0)(B) } p(dz, t_1) \, = \, \int_B p(dz,t_0) 
\;\;\;\; \mbox{ for all Borel subsets } B \subset \Gamma_n,  
\;\; \mbox{ and any } \;\; t_1 \ge t_0 \, .   
 \]
We consider only  probability measures, $p(dz,t) =
\rho(z,t)dz$, having densities $\rho$ that are smooth in $z$ and $t$,
because all trial densities in our reduced model have this regularity.
Then the propagation of probability by the phase flow is governed by
the Liouville equation
\begin{equation}  \label{liouville_eqn} 
\frac{\partial \rho}{\partial t} \,+\, L \rho \,=\,0 
\;\;\;\; \mbox{ in } \Gamma_n \times \R \, ,    
\end{equation}
in which we introduce the Liouville operator $L= \{ \cdot , H \}$.
Given a density $\rho(z,t_0)$ at an initial time $t_0$,
(\ref{liouville_eqn}) completely determines the density $\rho(z,t)$ at
any later time $t$, denoted formally by $\rho(\cdot, t)
= e^{- (t-t_0) L} \, \rho(\cdot, t_0)$.  The statistical mean of
any dynamical variable $F$ at time $t$ is given by
\[
\la F | \rho(t) \ra \,=\, \int_{\Gamma_n} F(z) \, \rho(z,t) \, dz 
\,=\, \int_{\Gamma_n}  F \circ \Phi(t-t_0) \, (z) \, \rho(z,t_0) \, dz \, ,
\]
where the first equality defines our notation for expectation.  
In particular, the evolution of the statistical average of the
relevant vector $\la A | \rho(t) \ra$ is determined by the exact
solution of (\ref{liouville_eqn}).  From the point of view of a
numerical computation, however, solving the Liouville equation
impractical because it requires the
simulation of an ensemble of exact trajectories for
the complex dynamics  (\ref{hamilton_eqn}).  
For this reason we resort to the following natural approximation procedure. 

We seek to approximate the  exact density $\rho(t)$
by a family of trial densities, $\rhotilde(t)$, that have a simple
and tractable analytical form.  
For reduction relative to a given resolved vector $A$, a natural choice  is 
supplied by the so-called quasi-equilibrium, or quasi-canonical, densities
\cite{Jaynes1,Jaynes2,Zubarev,Penrose} already displayed in (\ref{qe-density}).
A standard motivation for using   the family of densities 
(\ref{qe-density})  to construct a statistical closure
is that each member of the family maximizes information entropy subject to
the mean value of the resolved vector; that is, $\rhotilde$ solves
\[
\mbox{ maximize } \; S(\rho) = - \int_{\Gamma_n} \rho \, \log \frac{ \rho }{\rhoeq} \,
dz \;\;\;\;\;\; \mbox{ subject to } \; \int_{\Gamma_n} A \rho \, dz \,
= \, a \, , \;\; \int_{\Gamma_n} \rho \, dz \,=\,1.
\]
From the perspective of information theory \cite{CT,Kullback},
$\rhotilde(z,\lambda)$ is the least informative probability density
relative to the equilibrium density $\rhoeq$ that is
compatible with the macrostate vector
\begin{equation}   \label{macrostate}
a \,=\, \la A | \rhotilde \ra \, = \, \int_{\Gamma_n} A(z) \rhotilde(z;\lambda) \, dz \, .
\end{equation} 
The parameter vector $\lambda \in  \R^m$ then consists of the Lagrange
multipliers for the vector constraint (\ref{macrostate}), and there is a one-to-one
correspondence between $\lambda$ and $a$ given by
\begin{equation}   \label{qe-duality}
a \,=\, \frac{\partial \phi}{\partial \lambda} \, ,  \;\;\;\;\;\;
\lambda \,=\, - \frac{\partial s}{\partial a} \, , 
\end{equation}
where $s(a)=S(\rhotilde)$ denotes the entropy of the macrostate $a$.
This correspondence is a convex duality, and $- s(a)$ is the convex
conjugate function to $\phi(\lambda)$.

In this formulation we fix an equilibrium density $\rhoeq$ on $\Gamma_n$
and construct the family of quasi-equilibrium trial densities $\rhotilde$ 
relative to it.   An alternative formulation is to include the Hamiltonian $H$
among the resolved variables, by forming an augmented resolved
vector $A=(A_0, A_1, \ldots , A_m)$ with $A_0=H$.  Then the trial
densities $\rhotilde = \exp [\lambda_0 A_0 + \cdots + \lambda_m A_m -
\phi (\lambda)]$ have the augmented parameter vector $\lambda \in R^{m+1}$
and are respect to phase volume $dz$.    This alternative leads to a parallel
theory.    Our formulation in terms of a non-conserved resolved vector $A$ 
focuses the discussion of nonequilibrium concepts and facilitates the
derivation of the near-equilibrium theory via the linear response approximation.

In physical terms the quasi-equilibrium description establishes a
nonequilibrium statistical mechanics without memory, in the
sense that  $\rhotilde(z,\lambda(t))$ depends only on the
instantaneous macrostate $a(t)$, not on its history, $a(t')$ for $t' <
t$.   Such a memoryless description might be justified when there is 
 a wide separation between the time scale of evolution of the
resolved vector, $A$,  and the time scale of
conditional equilibration of the unresolved variables.  
In the limit of an infinite time-scale separation, an instantaneous statistical closure 
with respect to these  quasi-equilibrium densities is obtained by
imposing the $A$-moment of the Liouville equation, namely,
\[
\frac{d}{dt} \la  A(z) \, | \,  \rhotilde(z;\lambda(t))   \ra 
\;=\; \la  L A (z) \, | \, \rhotilde(z;\lambda(t)) \, \ra \, .   
\]
But the resulting statistical closure is exactly entropy conserving,
as the following straightforward calculation shows:
\[
\frac{d}{dt} S(\rhotilde)  \,=\, - \lambda^* \frac{d a} {dt} \, = \, 
  \int_{\Gamma_n} L  \, \lambda^*A  \exp[\lambda^*A - \phi(\lambda)] \, \rhoeq \,  dz \,= 0 \, .   
\]
Thus, the combination of a quasi-equilibrium ansatz and an
instantaneous moment closure results in a reversible reduced dynamics
\cite{Dougherty,GK}.  The inability of this adiabatic closure to capture the
dissipation, or entropy production, of the actual nonequilibrium
statistical behavior is a serious
defect, making it useful only for certain short-time phenomena.

In much previous work \cite{Zubarev,ZMR,LVR,Ramshaw,GK}, the remedy
for this defect has been to include memory effects into the relevant
probability densities by replacing $\lambda(t)^*A$ with time-weighted
averages of the dynamical variables, $\int_0^{\infty}
\lambda(t-\tau)^* \, A \circ \Phi(-\tau) \, w(\tau) \, d \tau$.
Closure is then obtained by taking instantaneous $A$-moments with
respect to these memory-dependent densities.  The resulting reduced
equations, however, depend upon the weighting function $w$, for which
there is no universal choice.   Moreover,  they involve convolutions over time,
and therefore require the evaluation of various correlations of $A$
and $L A = \{A,H\}$ over a range of time shifts $\tau$.  The statistical
closures derived in this manner are consequently difficult to justify
theoretically and expensive to implement computationally.

The new approach that we propose in the next section is fundamentally
different.  Rather than use memory-dependent densities and moment closure, 
we retain the 
quasi-equilibrium densities as natural trial densities in a parametric statistical model.   
With respect to this model 
we obtain closure by best-fitting  trajectories of trial densities to the Liouville equation
over the entire time interval from when the initial statistical state is specified
to when the evolved state is estimated.   

\section{Formulation of best-fit closure}   

 As outlined
in the Introduction, we quantify the lack-of-fit of the quasi-equilibrium trial
probability densities (\ref{qe-density})  to the Liouville equation
(\ref{liouville_eqn}) over an interval of time $t_0 \le t \le t_1$ by a cost functional 
\begin{equation}  \label{cost-functional}
\Sigma[\lambda;t_0,t_1] \,=\, \int_{t_0}^{t_1} 
\L (\lambda, \dot{\lambda}) \, dt \, .
\end{equation} 
In this formulation the cost functional is analogous to an action
integral for a Lagrangian $\L$, which is a function of $\lambda$ and
$\dot{\lambda} = d\lambda/dt$, and  the configuration space
is the  parameter space $\R^m$ for the statistical model ({\ref{qe-density}).    To construct
an appropriate (running) cost function $\L(\lambda,\lambdadot)$, we introduce the  residual
of the log-likelihood, $\log \rhotilde$, of the trial densities with respect to the Liouville operator:
\begin{equation}   \label{residual}
R(\cdot;\lambda,\lambdadot) \, = \, \left( \frac{\d}{\d t} + L \right) \log \rhotilde(\cdot;\lambda(t))
   \;=\;   \lambdadot(t)^* (A - a(\lambda(t))) + \lambda(t)^*LA  \, .   
\end{equation}    
There are two related expressions for this Liouville residual $R$ that reveal its significance
in the statistical closure problem.    

First, for any $z \in \Gamma_n$
and any smooth parameter path $\lambda(t)$, we examine the log-likelihood ratio 
between the exact density and the trial density after a short
interval of time $\Delta t$:
\[
\log \frac{e^{-(\Delta t) L}  \rhotilde(z;\lambda(t)) }{ \rhotilde(z;\lambda(t+\Delta t))}
\; = \;  - (\Delta t) R(z;\lambda(t),\lambdadot(t)) \; + \; O ( \, (\Delta t )^2 \, )   
     \;\;\;\;\; \mbox{ as } \;\; \Delta t \rightarrow 0 \, .   
\]
This expansion shows that,  to leading order locally in time, $-R(z;\lambda(t),\lambdadot(t))$  represents the information in the sample point $z$ for discriminating the exact density 
against the trial density  \cite{Kullback}.
By considering arbitrary smooth paths passing through a point $\lambda$ with a tangent vector $\lambdadot$,  we may consider $R(z;\lambda,\lambdadot)$ 
evaluated at $z$ as a function of $(\lambda,\lambdadot)$ in the tangent space
to the configuration space, and we may interpret it to be the local rate of information loss in
the sample point $z$ for the pair $(\lambda,\lambdadot)$.     
 
 Second, for any  dynamical
variable $F: \Gamma_n \rightarrow \R$, we evaluate the $F$-moment of the Liouville
equation with respect to the trial densities along a path $\lambda(t)$ and we obtain
the identities
\[
\frac{d}{dt} \la F \, |\,  \rhotilde(\lambda(t)) \ra - \la LF \, | \,   \rhotilde(\lambda(t)) \ra
\,=\, \la \,F \, | \, (  \d / \d t + L ) \rhotilde(\lambda(t)) \, \ra   
 \,=\,  \la FR \, |\,   \rhotilde(\lambda(t)) \ra \,  
\]
where the second equality follows directly from the definition (\ref{residual}).   
Thus we find that, while an exact solution $\rho(t)$ of the Liouville equation satisfies
the  identities, $d/dt \la F |\rho(t) \ra - \la LF | \rho(t) \ra = 0$, for all 
test functions $F$, a trial solution $\rhotilde$ produces
a departure that coincides with  the covariance between
$F$ and $R$.   This representation  of $R$ furnishes a natural linear
structure for analyzing the fit of the trial densities to the Liouville equation.

In light of these two interpretations of the Liouville residual $R$,  we proceed to measure the
dynamical lack-of-fit of the statistical model in terms it.    To do so, we consider the
components of $R$ in the resolved and unresolved subspaces.  At any configuration point
$\lambda \in \R^m$, let $P_{\lambda}$ denote the orthogonal projection of 
the Hilbert space $L^2(\Gamma_n, \rhotilde(\lambda))$  onto the span of the functions
 $A_1 - a_1(\lambda), \ldots , A_m-a_m(\lambda)$, 
and let $Q_{\lambda} = I - P_{\lambda}$ denote 
the complementary  projection; specifically,  for any $F \in L^2(\Gamma_n, \rhotilde(\lambda))$,  
\[
P_{\lambda} F = 
    \la F (A-a(\lambda))^* \, | \,  \rhotilde(\lambda) \ra \,C(\lambda)^{-1} ( A - a(\lambda))  \, . 
\]
where 
\begin{equation}  \label{fisher}
C(\lambda) = \la (A-a(\lambda)) (A-a(\lambda))^* | \rhotilde(\lambda) \ra  \, .   
\end{equation}   
We take the cost function in (\ref{cost-functional}) to be 
\begin{equation}  \label{lagrangian}
\L (\lambda, \dot{\lambda})  \,=\,  \frac{1}{2}   \la   (P_{\lambda} R)^2 \, | \, \rhotilde(\lambda) \ra
           + \frac{\epsilon^2}{2} \la (Q_{\lambda} R)^2 \, | \, \rhotilde(\lambda) \ra
\end{equation}
for a constant $\epsilon \in (0,1]$.   This lack-of-fit norm is of mean-squared type,
 but with an adjustable parameter
$\epsilon$ that controls the weight given to the unresolved component of the residual
versus the resolved component.   

By duality,  $\L(\lambda,\lambdadot)$ can be characterized in terms of test functions $F$ as follows:
\[
2 \L(\lambda,\lambdadot) \, = \, \max \{ \, \la F R | \rhotilde \ra^2 \, : \; 
      \la   (P_{\lambda} F)^2 \, | \, \rhotilde \ra
           + \epsilon^{-2} \la (Q_{\lambda} F)^2 \, | \, \rhotilde \ra \, \le 1, \; \;
                F \in L^2(\Gamma_n, \rhotilde) \,  \}    \, .
\]   
This dual form of the norm highlights the fact that $\L$ weights departures
in the unresolved dynamical variables  less than departures in the resolved variables,
depending on $\epsilon$.   In particular, when $\epsilon$ is small the constraint in 
this maximization gives preference to those $F$ which have a relatively small 
component in the unresolved subspace.   Thus, as the adjustable parameter
$\epsilon$ is decreased, the contribution of the unresolved variables to the
cost function $\L$ is correspondingly decreased;  and, in the limit as $\epsilon
\rightarrow 0$, it vanishes entirely.   In Section 7 we explain how 
$\epsilon$ is related to time scales and memory effects.     

Before presenting the optimization problem that defines our best-fit closure, we first
exhibit the Lagrangian $\L$ in a more explicit form.  The Liouville residual has
zero mean, $\la R | \rhotilde(\lambda) \ra=0$, and 
its orthogonal components are given by
\[
P_{\lambda} R = [\, \lambdadot - C(\lambda)^{-1} \la LA | \rhotilde(\lambda) \ra \, ]^* (A-a) \, , 
\;\;\;\;\;\;\;\;   Q_{\lambda} R =  \lambda^*(Q_{\lambda} LA) \, .
\]
The calculation of $P_{\lambda} R$ employs the  string of equations,
\[
\la LA \,|\, \rhotilde \ra = \la (LA) e^{\lambda^*A - \phi(\lambda)} \,|\, \rhoeq \ra = 
   - \la (A-a) L e^{\lambda^*A - \phi(\lambda)} \,|\, \rhoeq \ra   = 
       - \la (A-a) L (\lambda^*A) \,| \, \rhotilde \ra  \, 
\]
in which the anti-symmetry of the operator $L$ with respect to $\rhoeq$ is used.  
Hence the Lagrangian cost function can be written in the explicit form, 
\begin{equation}   \label{lagrangian-explicit}
\L (\lambda, \dot{\lambda})  \,=\, 
  \frac{1}{2} [\, \lambdadot - C(\lambda)^{-1} \la LA | \rhotilde(\lambda) \ra \, ]^*C(\lambda)
                       [\, \lambdadot - C(\lambda)^{-1} \la LA | \rhotilde(\lambda) \ra \, ]
     \;+\; \frac{\epsilon^2}{2} \lambda^*D(\lambda)\lambda \, , 
\end{equation}
where 
\begin{equation}  \label{D-matrix}
D(\lambda) = \la (Q_{\lambda} LA)(Q_{\lambda} LA^*) \, | \, \rhotilde(\lambda) \ra \, . 
\end{equation}   
By analogy to analytical mechanics, one may regard the first member in (\ref{lagrangian-explicit})
as the ``kinetic" term and the second member as the ``potential" term.      The kinetic
term is a quadratic form in the generalized velocities $\lambdadot$ with positive-definite
matrix $C(\lambda)$.   In fact, 
this matrix is the Fisher information matrix \cite{Kullback,CB}
for the exponential family (\ref{qe-density}), having components
\[
C_{ij}(\lambda) \, = \, 
  \la  \, \frac{\d \log \rhotilde}{\d \lambda_i}  \frac{\d \log \rhotilde}{\d \lambda_j} \, | 
               \, \rhotilde(\lambda) \, \ra \, .
\]
It defines a natural Riemannian metric, $ds^2 = \sum_{i,j} C_{ij}(\lambda) d\lambda^i d\lambda^j$,
 on the configuration space $\R^m$.   
 The potential term equals $\epsilon^2$ times the function
\begin{equation}   \label{closure-potential}
w(\lambda) \, = \,  \frac{1}{2} \lambda^*D(\lambda)\lambda \, 
    = \, \frac{1}{2} \la [Q_{\lambda} L \log \rhotilde ]^2 \, | \, \rhotilde(\lambda) \ra \, ,
\end{equation}
which we call the closure potential because it embodies the influence of the unresolved
variables on the resolved variables.    Of course, these mechanical analogies are not literal.
Indeed, even though we refer to  $\L$ as a ``Lagrangian," it has the units of a rate of entropy production, not of an energy, and  it is a sum of two positive-definite terms, not a difference.

For a given initial time $t_0$, we define the value function \cite{BH,FR}
\begin{equation}   \label{value}
v(\lambda_1,t_1) \, = \, 
\min_{\lambda(t)=\lambda_1} \; v_0(\lambda(t_0)) \, + \, \Sigma[\lambda;t_0,t_1]   \, ,
\end{equation} 
in which the minimization is over all regular paths $\lambda(t)$ on the
interval $t_0\le t \le t_1$ that terminate at an arbitrary (admissible) state $\lambda_1 \in \R^m$
 at time $t_1$.    The optimization problem in (\ref{value}) is the foundation of
our best-fit closure scheme.     The value function
 (\ref{value}) quantifies a total lack-of-fit of the statistical state $\lambda_1$
at time $t_1$ with respect to evolution from an 
incompletely specified statistical state at time $t_0$.      As
explained above,  the second member of the objective functional in (\ref{value})
is a dynamically intrinsic norm on the Liouville residual for a path of densities $\rhotilde(\lambda(t))$ 
over the time interval, $t_0 \le t \le t_1$. 
The first member of the objective functional determines the  initial value function, $v_0(\lambda)$, 
which is given data in the optimization problem.     In the special case
when initial density, $\rho_0$, is specified to be a quasi-equilibrium density,
 $\rhotilde(\cdot; \lambda_0)$, for a known $m$-vector $\lambda_0 $, 
 the initial value function $v_0$ is singular; namely, 
\[
v_0(\lambda) \; = \; \left\{ \begin{array}{cc}  +\infty  \;\;\;\; &  \lambda \neq \lambda_0 \\ 
                                                                   0 \;\;\;\;  &  \lambda =  \lambda_0 \end{array}  \right.
\]
In this case the first member in the objective functional in (\ref{value}) can be dropped
and the constraint $\lambda(t_0)=\lambda_0$ imposed instead.   
This singular initial condition can also be obtained as the limit of penalty functions
$ v_0 (\lambda) = b (\lambda - \lambda_0)^2 / 2$ as $ b \rightarrow +\infty  \, $.   In the
general case when there is some uncertainty in the initial statistical state, $\lambda(t_0)$ is free
and is penalized by the nonsingular initial value function $v_0$, which represents the 
degree of specification of the initial statistical state.    
 
It is a fundamental fact from the calculus of variations that such a value function
satisfies a corresponding Hamilton-Jacobi equation \cite{Lanczos,GF,Evans}.   
Since $\L$ is quadratic and  convex in $\lambdadot$, the required Legendre transformation
can be calculated explicitly.   The conjugate canonical variable is  
\begin{equation}  \label{mu}
\mu = \frac{\partial \L}{\partial \lambdadot} \;=\; C(\lambda) \lambdadot \,-\, 
           \la L A \, | \, \rhotilde(\lambda) \ra    \, , 
\end{equation}
 and the Hamiltonian associated with $\L$ is 
\begin{equation}   \label{hamiltonian}
\H (\lambda, \mu) \,=\,  \lambdadot^*\frac{\partial \L}{\partial \lambdadot} - \L
                  \,=\, \half \mu^* C^{-1} \mu \, + \,   \la L A \, | \,  \rhotilde(\lambda) \ra^* C^{-1} \mu
             - \epsilon^2 w(\lambda)       \, ,          
\end{equation}
where we recall the closure potential $w(\lambda)$ defined in (\ref{closure-potential}).  
The value function in (\ref{value}) therefore satisfies the Hamilton-Jacobi equation
\begin{equation}  \label{hj-equation}
\frac{\partial v }{ \partial t }  \, + \, \H (\lambda, \frac{\partial v }{ \partial \lambda}   ) \, = \, 0 \, ,
\;\;\;\; \mbox{ for } \; t > t_0 \, , \;\;\;\; \mbox{ with } \;\; v(\lambda,t_0) = v_0(\lambda)  \, . 
\end{equation}   

At this juncture of our development we have a complete formulation of the
desired model reduction scheme.   Given a resolved vector, $A$, 
and an equilibrium density, $\rhoeq$, the lack-of-fit Hamiltonian (\ref{hamiltonian})
is entirely determined up to a choice of the parameter $\epsilon \in (0,1]$, and 
the Hamilton-Jacobi equation (\ref{hj-equation}) propagates the value function $v=v(\lambda,t)$
forward in time from any suitable initial value function $v_0(\lambda)$.    We may view
this Hamilton-Jacobi equation as the  appropriate contraction of the Liouville
equation relative to our statistical model reduction in the following sense.    For
each instant of time $t>t_0$,   $v=v(\lambda,t)$ is a function on the statistical
parameter space  that may be conceptualized as a dynamical analogue of 
a minus-log-likelihood function \cite{CB,Kullback}.   The minimizer, $\lambdahat=\lambdahat(t)$,
of $v(\lambda,t)$ defines the best-fit estimate of the parameter $\lambda$ at time $t$,
and thus $\lambdahat$ is analogous to a maximum likelihood estimate.   
Moreover, the second-order behavior of $v(\lambda,t)$ 
in a neighborhood of $\lambdahat$ represents the confidence in, or
information content of,  the best-fit estimate. 
Thus, we may say that, just as the Liouville equation propagates the
ensemble probability on phase space, the Hamilton-Jacobi equation (\ref{hj-equation})
propagates the uncertainty in the reduced model, which is quantified by $v$.     
In this light, we discover that the desired reduced model closes 
at the level of the value function $v$ on configuration space, not at the level of the estimated
configurations themselves.   This property of our best-fit closure reflects the
fact that the value function governs both the best-fit estimate, at first order, and
its uncertainty, at second order.     

Summarizing, we define the best-fit parameter and macrostate by
\begin{equation}    \label{best-fit-parameter}
\lambdahat(t) \, = \, \arg \min_{\lambda} v(\lambda,t) \, ,    \;\;\;\;\; 
    \ahat(t) \,=\, \frac{\d \phi}{\d \lambda} (\lambdahat(t))  \;\;\;\;\;\; 
            \mbox{ for each } \;\; t \ge t_0 \, .
\end{equation}  
Accordingly, our best-fit closure is entirely determined by the solution of the 
 Hamilton-Jacobi equation (\ref{hj-equation}).   
 Moreover, this best-fit estimation of time-dependent, nonequilibrium statistical states 
 is valid far from equilibrium.    While this general result is a certainly satisfactory
from a theoretical viewpoint, it suffers from the fact that the nonlinear partial differential equation
(\ref{hj-equation}) is difficult to solve except when the number $m$ of resolved variables is very small.
For this reason, we now proceed to derive ordinary differential equations for $\lambdahat(t)$, 
or equivalently $\ahat(t)$, and investigate to what extent it is possible to circumvent solving the Hamilton-Jacobi equation.

\section{Derivation of closed reduced dynamics} 

We are primarily interested in the evolution of the best-fit 
parameter $\lambdahat(t)$,  and correspondingly
the best-fit macrostate $\ahat(t)$, since they determine
our estimates of the  expectations $\la B | \rhotilde(\lambdahat(t)) \ra$ 
of any dynamical variable $B$.    We therefore seek the system of ordinary differential
equations for $\lambdahat(t)$.  

By definition $\lambdahat(t)$ is the minimizer for $v(\lambda,t)$ for each $t > t_0$,
and consequently it satisfies 
\begin{equation}  \label{bf-equation}
 \hat{\mu}(t) \ = \, \frac{\partial v}{\partial \lambda}  (\lambdahat(t), t) \, = \, 0   \;\;\;\;\;\;\;\;
    \mbox{ for all } \; t > t_0 \, . 
\end{equation}
Differentiating this relation with respect to $t$, and invoking the Hamilton-Jacobi equation, 
we get
\begin{eqnarray*}
0  & = &  \frac{\partial^2 v}{\partial \lambda \partial \lambda^* } \cdot \frac{d \lambdahat}{dt}
 \, + \,  \frac{\partial^2 v}{\partial \lambda \partial t}     \\
    &= &  \frac{\partial^2 v}{\partial \lambda \partial \lambda^* } \cdot \frac{d \lambdahat}{dt}
  \, - \,  \frac{\partial}{\partial \lambda} 
           \H (\lambda, \frac{\partial v}{\partial \lambda} ) \, |_{\lambda=\lambdahat} \\
    & =&      \frac{\partial^2 v}{\partial \lambda \partial \lambda^* } \cdot \frac{d \lambdahat}{dt}
   \, - \,   \frac{\partial \H}{\partial \lambda} (\lambdahat, 0 ) 
   \, - \, \frac{\partial^2 v}{\partial \lambda \partial \lambda^* } \cdot \frac{\partial \H}{\partial \mu}  
              ( \lambdahat, 0 )    \, ,    
\end{eqnarray*}
where 
\[
 \frac{\partial^2 v}{\partial \lambda \partial \lambda^* } = 
     \left(  \frac{\partial^2 v}{\partial \lambda^i \partial \lambda^j }  \right)_{i,j=1, \ldots m}
\]     
denotes the  Hessian matrix of $v$ with respect to $\lambda$.  
The second and third terms in this final expression are calculated from $\H$ to be
\[
 \frac{\partial \H}{\partial \lambda} (\lambdahat, 0 )  \, = \, 
       - \epsilon^2  \frac{\partial w}{\partial \lambda} (\lambdahat ) \; , 
    \;\;\;\;\;\;\;\;
\frac{\partial \H}{\partial \mu} (\lambdahat, 0 ) \,=\, 
    C(\lambdahat)^{-1} \la \, L A  \, | \, \rhotilde(\lambdahat) \, \ra  \, .   
\]
Solving for $d \lambdahat/dt$, we arrive at the equation governing the reduced dynamics:
\begin{equation}   \label{closed-reduced-eqn} 
\frac{d \lambdahat}{dt} \; = \;  
   C(\lambdahat)^{-1} \la \,  LA \, | \, \rhotilde (\lambdahat) \, \ra 
    \; - \; \epsilon^2 \left[  \frac{\partial^2 v}{\partial \lambda \partial \lambda^* }(\lambdahat,t) \right]^{-1}
                  \frac{\partial w}{\partial \lambda} (\lambdahat ) \; .  
\end{equation} 

This is a first-order system of ordinary differential equations is closed in $\lambdahat$, but it involves
the value function $v(\lambda,t)$, which in turn is determined by the Hamilton-Jacobi
equation.   The first term on the right-hand side of (\ref{closed-reduced-eqn}) 
is exactly what is obtained
from the adiabatic closure when the moment condition $\la R (A - a) \, |\,  \rhotilde \ra = 0$
is imposed instantaneously at each time $t$; and thus the adiabatic
closure is recovered from the best-fit closure in the limit as  $\epsilon \rightarrow 0$.   
The dissipation, or irreversibility, in the reduced dynamics is produced by second
term on the right hand side of (\ref{closed-reduced-eqn}).     
The magnitude of this dissipation depends on the adjustable parameter $\epsilon$, but it 
does not necessarily scale like $\epsilon^2$, owing to the $\epsilon$-dependence of 
$v$ through the Hamilton-Jacobi equation.    
 
We now attempt to find a governing equation for the Hessian matrix, 
\begin{equation}  \label{hessian}
\Mhat(t) =   \frac{\partial^2 v}{\partial \lambda \partial \lambda^* } (\lambdahat(t),t) \, 
\end{equation}
since the inverse of this matrix enters into (\ref{closed-reduced-eqn}).  To do so, we
calculate the matrix of all second partial derivatives with respect to $\lambda$ 
of the Hamilton-Jacobi equation (\ref{hj-equation}), and thereby obtain
a partial differential equation for the matrix-valued function
\[
M(\lambda,t) =   \frac{\partial^2 v}{\partial \lambda \partial \lambda^* } (\lambda,t) \, .  
\]
The result is
\begin{equation}  \label{second-derivative-eqn}
0 \, = \,  \frac{\d M}{\d t}  \, + \,  
     \frac{\partial^2 \H}{\partial \lambda \partial \lambda^* } \, + \, 
     \frac{\partial^2 \H }{\partial \lambda \partial \mu^* }  M  \, + \, 
     M \frac{\partial^2 \H}{\partial \mu \partial \lambda^* }    \, + \, 
      M   \frac{\partial^2 \H}{\partial \mu \partial \mu^* }  M    \, + \, 
        \sum_{k=1}^m  \frac{\d \H} {\d \mu_k}  \frac{\d M}{\d \lambda_k}  \; . 
\end{equation}
In order to set $\lambda=\lambdahat$ and $\hat{\mu}=0$ in this matrix equation, 
we need to evaluate the various coefficients in it, using the expression 
(\ref{hamiltonian}) for $\H$.   We introduce the shorthand notation
\[
 f(\lambda)  \, = \,  C(\lambda)^{-1} \la LA \, | \, \rhotilde(\lambda) \ra \, , 
 \]
 which coincides with $d \lambda / dt$ under the adiabatic closure. 
 The required coefficients are 
 \[
\frac{\d \H}{\d \mu} (\lambdahat,0) = f(\lambdahat) \, , 
\]
\[
\frac{\partial^2 \H}{\partial \lambda \partial \lambda^* }(\lambdahat,0) = 
 -\epsilon^2 \frac{\d^2 w}{\d \lambda \d \lambda^*} (\lambdahat) \, ,     \;\;\;\;
\frac{\partial^2 \H}{\partial \mu \partial \mu^* }(\lambdahat,0) =
 C(\lambdahat)^{-1} \, ,   
\]
\[
 \frac{\partial^2 \H }{\partial \lambda \partial \mu^* }(\lambdahat,0) =  
         \frac{\d f} {\d \lambda}(\lambdahat) \, , \;\;\;\;\;\; 
 \frac{\partial^2 \H}{\partial \mu \partial \lambda^* }  = 
       \left[ \frac{\d f} {\d \lambda}(\lambdahat) \right]^* \, .    
\]
Substituting these identities into (\ref{second-derivative-eqn}) we obtain:
\[
0 \, = \, \frac{\d M}{\d t}(\lambdahat,t) - \epsilon^2 \frac{\d^2 w}{\d \lambda \d \lambda^*} 
        +  \frac{\d f} {\d \lambda} \Mhat 
        + \Mhat \left[  \frac{\d f} {\d \lambda} \right]^*   
         + \Mhat C^{-1} \Mhat  \,  +  \,  f \cdot \frac{\d M}{\d \lambda} (\lambdahat,t)   \, , 
\]
in which all the coefficient functions are evaluated at $\lambda = \lambdahat(t)$.
Now using the governing equation (\ref{closed-reduced-eqn}) for $\lambdahat$,
which can be written as, 
\[
\frac{d \lambdahat}{d t} \, = \, f(\lambdahat) \, - \,  
                 \epsilon^2 \Mhat^{-1} \frac{\d w}{\d \lambda} (\lambdahat)  \, , 
\]
we find that the matrix-valued function $\Mhat(t) = M(\lambdahat(t),t)$ satisfies 
\begin{equation}  \label{Mhat-eqn}
 \frac{d \Mhat}{d t} \, = \,  - \frac{\d f} {\d \lambda} \Mhat 
        - \Mhat \left[  \frac{\d f} {\d \lambda} \right]^*   
         - \Mhat C^{-1} \Mhat  \,+  \epsilon^2 \frac{\d^2 w}{\d \lambda \d \lambda^*} \, 
             -   \epsilon^2 \left[ \Mhat^{-1} \frac{\d w}{\d \lambda}  \right] \cdot 
                             \frac{\d M}{\d \lambda} (\lambdahat,t)  \, , 
\end{equation}  
in which again $\lambda = \lambdahat(t)$ throughout this equation.     

Our goal is to close the differential equation  (\ref{closed-reduced-eqn}) for
the best-fit parameter $\lambdahat$ by coupling it to a differential equation
for $\Mhat$.   But the last term in (\ref{Mhat-eqn}) involves the partial derivatives 
$\d M / \d \lambda$, and consequently it is not closed in $\Mhat$.     In fact, 
the  pair of first-order differential equations (\ref{closed-reduced-eqn}) and (\ref{Mhat-eqn})
constitute the first and second members of a closure hierarchy.    
In principle, we could generate a third member of the hierarchy by deriving a differential equation 
for $\d M / \d \lambda (\lambdahat, t)$, and so forth.   We will not pursue the hierarchy
of ordinary differential equations any further, since the best-fit estimation scheme closes 
elegantly at the level of the Hamilton-Jacobi equation for the value function.   

Nonetheless, it is important  to achieve an approximate closure scheme
that is more computationally tractable that solving the Hamilton-Jacobi equation
itself.    Perhaps the most natural approximation of this kind is to 
set $\d M / \d \lambda (\lambdahat,t)=0$ identically.   We call this the 
local quadratic approximation, because it amounts to replacing the solution
$v(\lambda,t)$ of the Hamilton-Jacobi equation by its second-order Taylor
expansion around $\lambdahat$;  that is, 
\[
v(\lambda,t) \, \approx \, v(\lambdahat,t)  \, + \, 
      \half (\lambda-\lambdahat(t))^*\Mhat(t) (\lambda - \lambdahat(t))
            \;\;\;\;   \mbox{ for }  \lambda \mbox{ near } \lambdahat(t)   \, . 
 \]     
Moreover, recalling that the value function $v$ may be  viewed as
a dynamical analogue of a minus-log-likelihood function, 
we may regard this local quadratic  approximation  as comparable to a quasi-Gaussian approximation.    Under this approximation, the last term in (\ref{Mhat-eqn}) disappears and consequently we obtain a closed first-order differential equation for $\Mhat$;  namely,
\begin{equation}  \label{closed-Mhat-eqn}
 \frac{d \Mhat}{d t} \, = \,  - \frac{\d f} {\d \lambda}(\lambdahat) \Mhat 
        - \Mhat \left[  \frac{\d f} {\d \lambda}(\lambdahat) \right]^*   
         - \Mhat C(\lambdahat)^{-1} \Mhat  \,+  
              \epsilon^2 \frac{\d^2 w}{\d \lambda \d \lambda^*}(\lambdahat) \, . 
\end{equation}
We thus arrive at the desired closed reduced equations in the pair $(\lambdahat,\Mhat)$.
The coupled pair of governing equations for the statistical closure
consists of a state equation (\ref{closed-reduced-eqn}) for $\lambdahat$ and
a matrix Riccati equation (\ref{closed-Mhat-eqn}) for $\Mhat$.   The
parameter vector $\lambdahat$ determines the best-fit macrostate, while
the matrix $\Mhat$ characterizes the uncertainty in the best-fit estimate,
up to the local quadratic approximation.    The initial condition for (\ref{closed-Mhat-eqn}) is 
\[
\Mhat(t_0) = \frac{\d^2 v_0}{\d \lambda \d \lambda^*} (\lambda_0) \, , \;\;\;\; \mbox{ where }
  \;   \lambda_0 = \arg \min_\lambda v_0(\lambda) \, . 
\] 
We assume that the initial value function $v_0$ is strictly convex, and hence
that $\Mhat(t_0)$ is positive-definite.        

We note that the matrix solution $\Mhat(t)$ of (\ref{closed-Mhat-eqn}) for $t \ge t_0$
is necessarily symmetric and positive-definite whenever the closure
potential $w$ is convex along the solution $\lambdahat$.    
This conclusion follows from
known properties of solutions of matrix Riccati equations,
which hold for (\ref{closed-Mhat-eqn}) provided that the symmetric matrices 
$C$ and $\d^2 w/ \d \lambda \d \lambda^*$ are both positive-semidefinite,
and the initial condition is positive-definite \cite{Kucera}.      
While the Fisher information matrix $C(\lambda)$ is positive-definite for arbitrary $\lambda$, 
the convexity of the closure potential is  ensured only near to equilibrium; indeed,  
\[
\frac{\d^2 w}{ \d \lambda \d \lambda^* }(\lambda) = 
       D(0) + O(|\lambda|)   \;\; \mbox{ as } \lambda \rightarrow 0 \,  ,
\]
and matrix $D(\lambda)$ defined in (\ref{D-matrix}) is positive-semidefinite by construction.
Far from equilibrium the closure potential could lose its convexity, however, and
the solution matrix $\Mhat$ could become singular.   
In that situation the closed reduced equations would no longer be well-defined.   
This behavior would signify a bifurcation in the best-fit parameter trajectory, which
we would interpret as a dynamic phase transition in the reduced model.

\section{Near-equilibrium approximation}

In this section we present the simpler and more explicit form of
best-fit quasi-equilibrium closure that results from applying a 
linear response approximation \cite{Chandler,Zwanzig}
 in a neighborhood of the statistical equilibrium state $\rhoeq$.    
As in the preceding developments, the equilibrium density can be
any invariant probability density for the underlying Hamiltonian
dynamics.   For instance, it can be a canonical Gibbs ensemble:  
\[
\rhoeq(z) \, = \, \exp( - \beta [ \, H(z) - \psi(\beta) \, ] ) \, ,   \;\;\;\;\;\;
\mbox{ with } \;\;   
  \psi(\beta) = \beta^{-1} \log \int_{\Gamma_n} \exp( -\beta H(z) ) \, dz  \, , 
\]
for some inverse temperature $\beta >0$.    We
denote the equilibrium mean of any dynamical variable or vector $F$ on
$\Gamma_n$ by $\la F \ra_{eq}$.   

Since the quasi-equilibrium densities (\ref{qe-density}) reduce to the 
equilibrium density $\rhoeq$ when $\lambda=0$, 
the near-equilibrium theory is the linearization of
the general theory around $\lambda=0$.   Throughout this analysis
we assume that the resolved vector $A$ is normalized
relative to equilibrium by $\la A \ra_{eq} = 0 $.  We make the usual linear
response approximation 
\begin{equation}    \label{linear-response} 
\rhotilde(z;\lambda)  \, \approx \, [\, 1 \,+\, \lambda^* A(z)\,] \, \rhoeq(z) \,  .
\end{equation}  
Under this approximation the Lagrangian $\L(\lambda,\lambdadot)$ 
defined in (\ref{lagrangian}) becomes
a quadratic form in the pair $(\lambda, \lambdadot)$:  
\begin{equation}   \label{ne-lagrangian}
 \L(\lambda,\lambdadot)  \, \approx \,
       \half [ \, \lambdadot - C^{-1} J \lambda \, ]^*  \, C  \, [ \lambdadot - C^{-1}  J \lambda \,  ] 
              \; + \;  \frac{ \epsilon^2}{2}  \, \lambda^* D \lambda \;  ,
\end{equation}
where
\begin{equation}  \label{ne-coefficient-matrices}
C = \la A  \, A^* \ra_{eq} \, , \;\;\;\;   J = \la (L A)  \, A^* \ra_{eq} \, , \;\;\;\;
D = \la (Q L A ) \, (Q L A^*)  \ra_{eq}  
\end{equation}
The major simplification that occurs in the near equilibrium regime is that the
coefficient matrices $C,J,D$ defining (\ref{ne-lagrangian})
 are constants determined by the equilibrium state $\rhoeq$.  
These matrices have the symmetry
properties: $C^*=C, \; J^*=-J, \; D^*= D$.  
The projections $P$ and $Q = I - P$ are also independent of $\lambda$, being
orthogonal operators on $L^2(\Gamma_n,\rho_{eq})$.  
The Legendre transformation (\ref{mu}) and (\ref{hamiltonian}) now yields
\[
\mu = C \lambdadot - J \lambda \, , \;\;\;\;\;\;\;\;\;\;
 \H (\lambda, \mu) = \half \mu^* C^{-1} \mu -  \lambda^*J C^{-1} \mu
                                          - \frac{\epsilon^2}{2} \lambda^* D \lambda \, ,  
\]
in which the conjugate canonical variable, $\mu$, is linear in the pair $(\lambda,\lambdadot)$,
and the Hamiltonian, $H(\lambda,\mu)$, is a quadratic form.   The closure
potential $w$ is the quadratic form,
\begin{equation}  \label{ne-closure-potential}
w(\lambda)  \, = \, \half \lambda^* D \lambda \, , \;\;\;\; \mbox{ where } \;\; 
                 D = \la (LA)(LA^*) \ra_{eq} + J C^{-1} J
\end{equation} 

Since the Hamiltonian is exactly quadratic, the Hamilton-Jacobi equation (\ref{hj-equation})
admits a solution that is a quadratic form in $\lambda$:
\[
v(\lambda,t) = \hat{v}(t)  \, + \, 
       \half (\lambda-\lambdahat(t))^* \Mhat(t) (\lambda -\lambdahat(t))                                       
\]
for some $m \times m$ symmetric matrix $ \Mhat(t)$, $m$-vector $\lambdahat(t)$
and  scalar $\hat{v}(t)$.    Substitution of this ansatz into the Hamilton-Jacobi
equation produces equations
corresponding to the quadratic, linear and constant terms in $\lambda$.  Namely,
\[
0 = \frac{d \Mhat}{dt} + \Mhat   C^{-1} \Mhat - J C^{-1} \Mhat + 
                          \Mhat C^{-1} J - \epsilon^2 D    \,  ,                                
\]
\[
0 = -\Mhat \frac{d \lambdahat}{dt} + \Mhat C^{-1} J \lambdahat - \epsilon^2 D \lambdahat \, , 
\]
\[
0 = \frac{d \hat{v}}{dt} - \frac{\epsilon^2}{2} \lambdahat^* D \lambdahat  \, . 
\]
The first of these equations is a matrix Riccati equation for $\Mhat$.   It is supplemented
by the initial condition $\Mhat(t_0) = M_0 $ for a given positive-definite,
symmetric matrix $ M_0 $.     The second of these equations 
determines the best-fit vector $\lambdahat(t)$, 
and its initial condition is $\lambdahat(t_0) = \lambda_0$.   
The third equation merely generates the additive constant
$\hat{v}(t)$, and its initial condition may be set to zero, $\hat{v}(t_0) = 0 $.  

It is transparent that the near-equilibrium approximation produces a best-fit
closure theory for which the local quadratic approximation discussed in the
preceding section holds globally.  The structure of the governing equations
for the parameter vector $\lambdahat$ and the matrix $\Mhat$ is the same
as in the general case, given in the preceding section, 
except that the coefficient matrices are constant and
the Riccati equation decouples from the state equation.   It is worthwhile to
summarize the closed reduced equations governing near-equilibrium:
\begin{eqnarray}   \label{ne-closed-reduced-eqns}
 \frac{d \lambdahat}{dt} &=&  [ \, C^{-1} J  - \epsilon^2 \Mhat^{-1} D \, ] \lambdahat  \\
 \frac{d \Mhat}{dt} &=&   J C^{-1} \Mhat -
                          \Mhat C^{-1} J -  \Mhat   C^{-1} \Mhat + \epsilon^2 D   \nonumber
\end{eqnarray}

The positive-definite, symmetric matrix $\Mhat(t)$ has the interpretation as 
the confidence, or information, in the best-fit estimate $\lambdahat(t)$ at each time $t$.  
In particular, if all the eigenvalues of $\Mhat(t)$ are large, then the value function
$v(\lambda,t)$ increases rapidly away from its minimum point at $\lambdahat$,
meaning that the best-fit estimate is sharp with respect to the lack-of-fit
norm for the Liouville equation.    In general, the eigenvalues and associated
eigenvectors of $\Mhat(t)$ characterize the multivariate sensitivity of the
best-fit estimate,  and their evolution in time quantifies 
the propagation of uncertainty in the reduced model.      The
inverse matrix, $\Ghat(t) = \Mhat(t)^{-1}$, furnishes an alternative characterization
of uncertainty in the sense that it is comparable to a covariance matrix
for the parameter vector $\lambda$.    Moreover, as a straightforward calculation
shows, the near-equilibrium closed reduced equations (\ref{ne-closed-reduced-eqns})
have the following equivalent form in terms of $\lambdahat$ and $\Ghat$:
\begin{eqnarray}   \label{Ghat-eqns}
 \frac{d \lambdahat}{dt} &=&  [ \, C^{-1} J  - \epsilon^2 \Ghat D \, ] \lambdahat  \\
 \frac{d \Ghat}{dt} &=&  
                 C^{-1} J \Ghat -  \Ghat J C^{-1}  -  \epsilon^2  \Ghat D \Ghat +  C^{-1}  \, . 
                           \nonumber
\end{eqnarray}
This form has the attractive feature that matrix inversion is eliminated
from the closed reduced dynamics.   Moreover, the fully specified initial condition,
$\lambda(0)=\lambda_0$ with improper 
$v_0$ or $M_0$, is readily handled by the homogeneous initial condition $\Ghat(0)=0$ 
on the Riccati equation in (\ref{Ghat-eqns}).   

Finally, 
the pair of equations (\ref{Ghat-eqns}) has a strong resemblance to the equations for a 
Kalman-Bucy filter \cite{BH,Davis,FR}.  
That is, we may view the pair of equations (\ref{Ghat-eqns}) 
as a state estimate equation for $\lambdahat$ coupled to a variance equation for $\Ghat$.  
This similarity might be expected from the
fact that our best-fit closure scheme is derived from a dynamic optimization principle,
as are filtering algorithms.    However, our problem is not a standard filtering problem, 
because we are concerned with fitting a statistical model to underresolved, 
deterministic dynamics rather than blending a stochastic dynamics with noisy measurements.   
Nonetheless, we can put our closure problem into a filtering format by regarding the linear
mapping, $\lambda \mapsto Q L (\lambda^*A)$, which takes a state $\lambda$ into the
unresolved component of the Liouville residual,  as the measurement process.   In this 
formal analogy the measured data is the zero function identically in time, 
so that the measurement error corresponds to the $L^2$-norm of $ Q L (\lambda^*A)$.
The defining minimization principle for our 
best-fit closure is thereby identified with a filtering principle for a squared-norm that blends 
the resolved and unresolved components of the Liouville residual using a weight factor $\epsilon^2$.  
This interpretation may have some utility beyond providing a conceptual connection.  Namely,
it suggests that our best-fit closure scheme could be combined naturally with actual continuous measurements,  and in this way the estimation of the resolved variables could be 
continually improved by a hybrid of closure and filtering.  But we will not pursue this
line of development in the present paper.

\section{ Simplification under time reversal symmetry}

Before discussing the role of the adjustable parameter $\epsilon$,  
which scales the dissipative effects  in the best-fit closure, it is
convenient first to display the simplified closed reduced equations
that result from symmetry with respect to time reversal.   Accordingly, 
we consider the special case in which each of the resolved
variables, $A_k \; (k=1, \ldots , m)\, , \;$ composing $A$ is even under time
reversal.  In this case, and in the case when $m=1$ and $A=A_1$ is an
arbitrary scalar variable, the near equilibrium equations can be 
solved explicitly and the thermodynamic properties of its
solutions can be derived easily.  Moreover, this case pertains to reduced
models that are purely dissipative, a physically important 
situation \cite{dGM}.

Let $\Theta : \Gamma_n \rightarrow \Gamma_n $ denote the time reversal
operator defined by $\Theta(q,p) = (q,-p)$ for $z=(q,p) \in \Gamma_n$.  
A dynamical variable $F$ is even under time reversal if
$F \circ \Theta = F$ on $\Gamma_n$.  We assume that the Hamiltonian
and $\rhoeq$ are even.  Then the phase flow $\Phi(t)$ on $\Gamma_n$
satisfies $\Phi(t) = \Theta \circ \Phi(-t) \circ \Theta$.  For any even
dynamical variable, $F$, it follows that $F \circ \Phi(t) = F \circ
\Phi(-t) \circ \Theta$.

For a  resolved vector, $A = (A_1, \ldots , A_m)$,  composed of even
component variables $A_k$,  the matrix $J= \la (LA) A^* \ra_{eq}$ vanishes.  
This result follows directly from the fact that $LA$ is odd under time reversal,
which is easily seen from the identities:
\begin{eqnarray*}  
(LA) \circ \Theta &=&  
\lim_{\Delta t \rightarrow 0} 
   \frac{A \circ \Phi(  \Delta t) \circ \Theta- A \circ \Theta}{\Delta t}  \\ \nonumber 
 &=&  \lim_{\Delta t \rightarrow 0} \frac{  A \circ \Phi(- \Delta t)  - A  }{\Delta t}            
  \; = \;  - (LA)   \, .  
\end{eqnarray*}
That  $J=0$ for any scalar resolved variable, $A \in \R^1$, is immediate
from the  skew-symmetry of $J$.   

In the case of even resolved variables, the closed reduced equations (\ref{Ghat-eqns}) 
have a scaling structure with respect to the adjustable parameter $\epsilon$.  
In terms of the rescaled time $\tau = \epsilon (t-t_0)$, we introduce the
fundamental solution matrix, $ \Psihat(\tau)$, for the state
equation in (\ref{Ghat-eqns}) expressed as an equation for  
the macrostate $\ahat(t) = C \lambdahat(t)$:
\[
\frac{d \ahat}{d t} \,  = \, -  \epsilon^2 C \Ghat(t;\epsilon) D C^{-1} \ahat
\]
Representing the solution, 
$\ahat(t) = \Psihat(\epsilon t) a_0 $, in terms of its initial state $a_0$, and  rewriting the 
solution of the Riccati equations in (\ref{Ghat-eqns}) in terms of the 
rescaled matrix $\Khat(\tau) = \epsilon C \Ghat(\tau/\epsilon;\epsilon) D$,  
we find that (\ref{Ghat-eqns}) is equivalent to the following pair of matrix equations:  
\begin{equation}  \label{even-ne-eqns}
\frac{d \Psihat}{d \tau}  =  -  \Khat  C^{-1} \Psihat \, ,  \;\;\;\;\;\;\;\; 
       \frac{d \Khat}{d\tau}  =  D - \Khat C^{-1} \Khat  \, .   
\end{equation}
These equations are supplemented by the initial conditions,
  $\Psihat(0)=I, \; \Khat(0) = 0 $,
which are appropriate to the situation when the arbitrary initial macrostate
$a_0 \in \R^m$ is completely specified; incomplete specification of the
initial statistical state changes  $  \Khat(0)$.     

The pair of matrix equations (\ref{even-ne-eqns}) 
can be simultaneously diagonalized and hence solved explicitly. 
Let $W$ be a matrix of normalized eigenvectors of $D$ relative to
$C$, and let $\Delta$ be the corresponding diagonal matrix of
eigenvalues, all of which are real and nonnegative.  The nonsingular
matrix $W$ diagonalizes $D$ relative to $C$, meaning that
\[
 W^* D W = \Delta  \;\;\;\;\;\;\;\;   \mbox{ and }  \;\;\;\;\;\;\;\;   W^* C W = I \, .  
\]
Making the substitutions,
$\bar{\Psi}(\tau)  = W^* \Psihat(\tau) (W^*)^{-1} $ and $ \bar{K} (\tau) = W^* \Khat(\tau) W$,
in (\ref{even-ne-eqns}),  we get the diagonalized matrix equations,  
\[
\frac{d \bar{\Psi}}{d \tau} = -  \bar{ K} \bar{\Psi} \, , \;\;\;\;\;\;\;\;
\frac{d \bar{ K}}{d \tau} =  \Delta -  \bar{K}^2 \, .   
\]
Their solutions are elementary: $\bar{\Psi}(\tau) = \mbox{sech} ( \sqrt{\Delta} \, \tau )$ and 
$\bar{K}(\tau) = \sqrt{\Delta} \tanh  ( \sqrt{\Delta} \, \tau )  $, 
where $\sqrt{\Delta}$ denotes the nonnegative square root of the
nonnegative diagonal matrix $\Delta$, and the hyperbolic functions
act in the obvious way.   
Inverting the transformation, we obtain the desired solutions of (\ref{even-ne-eqns});
namely,
\begin{equation}  \label{special-matrix-solns}
\Psihat(\tau) = (W^*)^{-1} \, \mbox{sech} ( \sqrt{\Delta} \, \tau ) \, W^* \, , \;\;\;\;\;\;\;\;
\Khat(\tau) = (W^*)^{-1} \, \sqrt{\Delta } \tanh ( \sqrt{\Delta} \, \tau ) \, W^{-1} 
\end{equation}

Returning to the closed reduced equations (\ref{Ghat-eqns}) expressed
in unscaled time $t$, we now have a relaxation equation for the best-fit
macrostate with an explicit, scaled, time-dependent coefficient matrix:
\begin{equation}   \label{scaled-even-eqn}
\frac{d \ahat}{dt} \, =\, - \epsilon \Khat (\epsilon (t-t_0)) \, \lambdahat(t) \, . 
\end{equation}
The format of (\ref{scaled-even-eqn}) is reminiscent of the linear transport
equations of phenomenological nonequilibrium thermodynamics
\cite{dGM,Keizer,Ottinger}.  In that setting, a separation of time scales
between the evolution of the macrostate and the fluctuations of the
microstate is assumed, and linear relaxation equations are posited to
relate fluxes to forces.  In our notation the thermodynamic forces are
$-\lambdahat$, while the thermodynamic fluxes are $d \ahat /dt$.  
In the well-known Onsager theory of near-equilibrium relaxation, the
matrix of transport coefficients is usually denoted by $L$, and
is constant in time, so that the phenomenological equations are
\[
\frac{d \ahat}{dt} \,=\, - L \lambdahat  \, ,  \;\;\;\;\;\;\;\;
 \mbox{ with }  \;\;\;\;    \ahat (t) = C \lambdahat (t) \, .
\]
For a resolved vector $A$ that is even under time reversal,
the celebrated reciprocity relations imply that $L$ must be
symmetric.  The entropy production is then $d \hat{s}/dt = \lambdahat^* L
\lambdahat$, and this expression is invoked to imply that $L$ must be
positive-definite.   These classical results rely on
 a number of assumptions concerning the format of the macroscopic
transport equations and the statistical properties of the microscopic
fluctuations, namely the Onsager regression hypothesis.  

By contrast,
our relaxation equation (\ref{scaled-even-eqn}) has a time-dependent transport matrix,
$ \epsilon \Khat (\epsilon (t-t_0))$,  that is derived
from the underlying dynamics via the best-fit model reduction,
up to the adjustable parameter $\epsilon$. 
Moreover, the best-fit transport matrix is necessarily positive-definite and
symmetric by virtue of the Riccati equation \cite{Kucera}.  Thus, our best-fit reduced
dynamics shares the key qualitative properties of phenomenological
irreversible thermodynamics:  for the near-equilibrium, even-variable
regime, it possesses positive entropy production and reciprocity relations.
Furthermore,  the derivation of the relaxation equation
(\ref{scaled-even-eqn}) from the Liouville equaiton does not require an extreme separation
of time scales.   Indeed, the time dependence of $\Khat$ implies that
there is a plateau effect  during which  $\Khat(\tau)$
increases from $0$ to its  asymptotic limit,
$\Khat_{\infty}$, which is determined by $\Khat_{\infty} C^{-1} \Khat_{\infty} = D$.  
This plateau effect is regulated by the matrices $C$ and $D$.   
More precisely, there are $m$ plateau time scales, $\tau_1, \ldots , \tau_m$,
defined by $ \mbox{diag } \{ 1/\tau_1, \ldots , 1/\tau_m\} = \sqrt{\Delta} $,
which give the rates at which the eigenvectors of $\Khat(\tau)$
equilibriate.     Thus, we see that the scaled time variable
$\tau = \epsilon (t-t_0)$  applies to the plateau effect, while the
original time variable $t$ pertains to the relaxation.

\section{Irreversibility and the parameter $\epsilon$}   

 In this section we turn our attention to the role that the adjustable parameter 
 $\epsilon$ plays in the best-fit approach to model reduction. 

Broadly speaking,  a reduced model for 
a complex dynamics with finitely-many degrees of freedom
may be expected to exhibit three distinct time scales:  (1) the
relaxation time scale, $T_r$, over which the macroscopic resolved
variables decay (and possibly oscillate); (2) a plateau time, $T_p$,
over which fluctuations in the unresolved variables  influence the resolved variables;
and (3) a memory time, $T_m$, over which the unresolved fluctuations
decorrelate.    Moreover, these time scales generically have the
ordering:  $ T_r >  T_p > T_m$.   In fact, a strong separation of
time scales,  $ T_r \gg  T_p \gg T_m$ is necessary to justify a 
Markovian stochastic model of the resolved variables.   
In the absence of such an extreme separation of time scales, 
the reduced dynamics is described by well-known generalized transport
equations that are derived by the Mori-Zwanzig projection method \cite{Zwanzig}.
This formal identity has recently been employed as the starting point
for statistical closure for underresolved Hamiltonian dynamics \cite{CHK1,CHK2,GKS}.
This work follows a line of development in nonequilibrium statistical
mechanics that have been pursued by various investigators in the past
\cite{Balescu,Prigogine_etal,Robertson,Ramshaw,ZMR,LVR}.   While our
approach is fundamentally different than these works, the generalized 
transport equations furnish a means of interpretation of our adjustable
parameter $\epsilon$ and an explanation of how $\epsilon$ is related 
to the relevant times scales in the statistical closure.  

We focus on the near-equilibrium of our theory and of the projection
operator identity.   To within the linear response approximation we are
given an initial ensemble 
$\rho_0 = \exp[\, \lambda_0^*A - \phi(\lambda_0) \, ] \, \rhoeq
\approx [ \, 1 +    \lambda_0^*A \, ] \rhoeq$, 
and we are interested in the evolution
of the ensemble-averaged macrostate 
$a_{ex}(t) = \la e^{(t-t_0)L}A \, | \,  \rho_0 \ra
\approx \la \, ( e^{(t-t_0)L}A ) \, A^*  \,  \ra_{eq} \, \lambda_0$.      
The following Mori-Zwanzig formula, which is an exact consequence of
the Liouville equation up to the near-equilibrium approximations,  states that
\begin{equation}   \label{mz}
\frac{d a_{ex}} { dt} \, = \, J C^{-1} a_{ex} \,  
  - \int_{t_0}^t Z(t-t') \, C^{-1} a_{ex}(t') \, dt' \, ,    
 \end{equation}   
 where $ Z(\theta) = \la [ \, e^{ \theta Q L }(Q L A)]  (Q L A^*)  \, \ra_{eq}$.  
The central difficulty faced when implementing this formula in practice is to
find tractable approximations to the exact memory kernel $Z(\theta)$, which 
involves the complementary orthogonal propagator, $e^{\theta QL}$, in 
memory time $\theta$.   

To relate this formula  to our best-fit closure,
let us attach a quasi-equilibrium density to the exact macrostate, $a_{ex}(t)$,
at each instant of time.   That is, we define the corresponding parameter vector
$\lambda_{ex}(t) = C^{-1} a_{ex}(t)$
and density $\rho_{ex}(t) = \rhotilde(\lambda_{ex}(t))$.   Of course, the goal
of our closure scheme is to 
approximate the exact state $\lambda_{ex}(t)$ by the estimated state $\lambdahat(t)$,
which is computed without evaluation of the memory kernel $Z(\theta)$ or
equivalent quantities.    For the purposes of interpreting $\epsilon$, we 
examine the entropy, $s_{ex} = \phi(\lambda_{ex}) - \lambda_{ex}^*a_{ex}$, 
of the quasi-equilibrium states attached to the exact states and calculate the 
associated entropy production.    From (\ref{mz}) we find that 
\begin{equation}    \label{entropy-prod-exact}
\frac{d s_{ex} } {dt} \, = \, \int_{t_0}^t \lambda_{ex}(t)^* Z(t-t') \lambda_{ex}(t') \, dt' 
\end{equation}
Correspondingly, the entropy production for our best-fit closure in the near-equilibrium
regime is given by 
\begin{equation}   \label{entropy-prod-model}
\frac{d \hat{s} } {dt} \, = \,  \epsilon^2 \lambdahat(t)^* C \Ghat (t; \epsilon) D \lambdahat(t) \, , 
\end{equation}   
where $\Ghat$ is determined by the Riccati equation in (\ref{Ghat-eqns}).   Roughly
speaking, the free parameter $\epsilon$ should be adjusted so that the
entropy production in (\ref{entropy-prod-model}) approximately matches that 
in  (\ref{entropy-prod-exact}).   Since the memory kernel is not accessible in the
best-fit reduced model, this adjustment is an empirical tuning \cite{PT}.   

To proceed further with the heuristic analysis, let us suppose that the resolved
variable $A$ is a scalar.   Then, $J=0$, and $C,D>0$.  
Recalling the discussion in the preceding section,  the plateau time is $T_p = \sqrt{D/C}$.   
We write the memory kernel in the form, $Z(\theta) = D \zeta (\theta / T_m)$, for a
dimensionless function $\zeta(u)$  with $\zeta(0) =1$ and 
$\lim_{u \rightarrow \infty} \zeta(u) = 0$; in other words, $Z$ decays with a
characteristic memory time scale $T_m$.   Then assuming that the relaxation
time scale, $T_r$, is large compared to $T_m$, we have
\[
\frac{d s_{ex} } {dt} \, = \, \frac{C }{T_p^2} 
      \int_{t_0}^t \lambda_{ex}(t)\,  \zeta(\frac{t-t'}{T_m}) \, \lambda_{ex}(t') \, dt'  
      \,  \approx  \,  \frac{C \, T_m}{T_p^2} \,   \lambda_{ex}(t)^2 
      \int_{0}^{\infty} \zeta(u) \, du \,  .
\]    
On the other hand, in this scalar case the best-fit closure gives the approximation
\[
\frac{d \hat{s} } {dt} \, \approx \,  \frac{C \epsilon}{T_p} \, \lambdahat(t)^2 \, , 
\]
supposing that the $T_r$ is large compared to $T_p$, and hence replacing
$\Ghat(t;\epsilon)$ by its asymptotic value $\Ghat_{\infty} = T_p/C \epsilon $.   
 These approximate expressions for the entropy production agree (up to some absolute
constants) provided that $\epsilon \sim T_m/T_p$.   Thus, we conclude  that $\epsilon$
effectively sets the memory time scale, which is not evaluated in the best-fit
closure scheme, relative to the plateau time, which is known.    Moreover,
a similar analysis applied to the relaxation equation for $\ahat$ in (\ref{Ghat-eqns})
shows that its decay rate is approximately $\epsilon^2 \Ghat_{\infty} D = \epsilon/T_p$.
Consequently, we obtain the complementary result that $\epsilon \sim T_p/T_r$.
The fact that $ T_p/T_r   \sim T_m/T_p$ is already implied by (\ref{mz}) when
there is a separation of time scales.    

In essence, 
the single adjustable parameter $\epsilon$ represents the minimal amount
of extra information about memory effects that must be included into the best-fit closure scheme.
Since $\epsilon$ enters into the lack-of-fit Lagrangian as a scale factor for
 the closure potential $w$ that determines the irreversible
terms in the closed reduced equations, it is manifest that $\epsilon$  regulates the magnitude of dissipation in the reduced model.    As the discussion in this section shows, the appropriate value
of $\epsilon$ is related to a common ratio of the characteristic time scales in the model
reduction problem.    But, from the point of view of implemented computations,  these
rough estimates do not suffice to determine $\epsilon$ quantatively.
Rather, $\epsilon$ must be estimated from some observations 
 or simulations that are not part of the best-fit closure itself.     

Our presentation throughout this paper has been restricted to a single adjustable
parameter $\epsilon$, and the heuristic analysis given above applies only to a
single scalar resolved variable $A$.  In a companion paper \cite{PT}, we implement the 
model reduction in this form and investigate quantitatively how well the statistical
closure scheme approximates fully resolved statistical solutions.   In such
investigations, and in other potential applications, the single parameter theory
can be expected to perform well when it is possible to identify a memory time
and a plateau time, at least approximately, that pertain to all the resolved variables.
In general, when multiple time scales operate,  a better approximation might be
obtained by a more intricate construction of the lack-of-fit functional.  Specifically,
we could replace the Lagrangian (\ref{lagrangian}) by 
\[
 \L (\lambda, \lambdadot) \, = \, \half \la (P_{\lambda} R)^2 \, | \, \rhotilde(\lambda) \ra
                 \, + \,  \half \la (E_{\lambda}Q_{\lambda} R)^2 \, | \, \rhotilde(\lambda) \ra  \, , 
\]
in which $E_{\lambda}$ is a self-adjoint, positive operator on 
$L^2(\Gamma_n,\rhotilde(\lambda)$, possibly depending on $\lambda$, with operator norm,  
$\|  E_{\lambda} \|  \le 1$.     Throughout the present paper
$E_{\lambda} = \epsilon I$.  By designing other operators $E_{\lambda}$
the resulting closure potential $w$ might more faithfully represent the
influence of the unresolved fluctuations on the resolved variables.   The best-fit
strategy would carry over to this generalization, but its practical implementation 
 would require  multiple tuned parameters. 

\section{Discussion and conclusions}

The methodology presented in this paper offers a new approach to the general
problem of constructing a reduced statistical-dynamical description of a complex
Hamiltonian system.   Most approaches to problems of this kind
in nonequilibrium statistical mechanics interpose a stochastic model 
between the given microscopic dynamics and the derived macroscopic dynamics,
while our approach deduces an irreversible macrodynamics for the selected set of 
resolved variables directly from the deterministic and reversible microdynamics.
Our method is based on an optimization principle for parameterized paths of trial probability
densities on phase space:
paths that achieve the least residual with respect to the Liouville equation
determine the evolution of the estimated macrostates.        
This approach provides an
information-theoretic meaning to the statistical closure, since  the best-fit paths 
minimize a cost functional that measures the rate of
information loss incurred by the model reduction.  Moreover, it leads to
a closure theory having a form analogous to
optimal filtering theory.  That is,   the best-fit closure blends unresolved 
microdynamics into the reduced equations for the resolved macrostate via
a matrix of transport coefficients in much the same way that the Kalman-Bucy filter assimilates 
continuous measurements into a stochastic state equation via a gain matrix.  
By formulating the prediction of underresolved dynamics as optimal estimation, 
the best-fit approach furnishes a new perspective 
on statistical closure of  complex, chaotic, or turbulent dynamics.  

The value function for the defining optimization principle is a key ingredient
in the best-fit closure theory.    It has the units of entropy production, it represents the optimal cost of reduction with respect to the selected resolved variables, and it solves the Hamilton-Jacobi equation associated with the cost function.    
The best-fit macrostate  corresponds to the evolving minimizer of the value function, while the uncertainty of the best-fit estimate is measured by the Hessian matrix of the value function at 
the minimizer.      We may say that our Hamilton-Jacobi equation plays a role analogous to that of the Fokker-Planck equation for a stochastic model whose resolved variables are described by a Langevin dynamics.       One conclusion of our work is that
irreversible behavior can be derived from an optimization principle and its related Hamilton-Jacobi theory, without explicitly introducing diffusive processes.     
 
Criteria for the choice of the resolved variables are not considered in this paper, but  
this is an important aspect of any model reduction strategy.  
In some physical problems it may be evident that certain
variables offer a thermodynamic, or kinetic, description, and thus  are
natural resolved variables.   The separation
of time scales between resolved and unresolved motions often informs this choice.    
More generally, though, there may be latitude in the selection of the resolved variables,
and the design of a good model reduction may require examining a variety 
of sets of resolved variables.    In the best-fit closure, a preferred choice would be one
for which the uncertainty of the estimated macrostate, as measured by the
second-order behavior of the value function,  is relatively small.  
In principle, therefore,  the theory itself
offers a method of distinguishing the predictive power of different sets of resolved 
variables.    We leave the issue of designing reduced descriptions, however,
 to future investigations of particular problems with the best-fit methodology.   

The framework for statistical model reduction that we propose in this work is
intended to be implemented on complex, multi-scale dynamical problems.   
To this end, the  local quadratic approximation of the
value function in the far-from-equilibrium regime and the related simplification in the
near-to-equilibrium regime are developed so that the numerical solution of the Hamilton-Jacobi equation itself is avoided.   Under these approximations the best-fit closure theory
admits a computational effective implementation, since sophisticated and powerful numerical methods exist for the integration of Riccati matrix differential equations. 
The best-fit approach does not require the computation
of memory kernels, which arise in approaches through Mori-Zwanzig projection methods
and necessitate either simulations of cumbersome propagators or analytical
approximations available only in limiting situations.    Instead, the tunable parameter
$\epsilon$  introduces a single time-scale separation into the best-fit reduction strategy.    
We recognize, though, that extensions of our basic
methodology would likely involve more elaborate trial densities and more detailed
 lack-of-fit norms, which would include multiple time-scale parameters.

\section{Acknowlegments}

In the course of this work, the authors benefited from conversations
with R.~S. Ellis, M. Katsoulakis,  
R. Kleeman, A.~J. Majda, F. Theil and S. Nazarenko.   
Some of this research was conducted during B.T.'s sabbatical visit to the
University of Warwick, which was partly supported by a international
short visit fellowship from the Royal Society.  B.T.'s work was funded
by the National Science Foundation under grants DMS-0207064 and
DMS-0604071.  P.P.'s research was partly sponsored by the Office of
Advanced Scientific Computing Research, U.S. Department of Energy. The
work was performed at the Oak Ridge National Laboratory, which is
managed by UT-Battelle, LLC under Contract No.  De-AC05-00OR22725.

\vspace{1cm}

\noi
{\bf Contact Information: }  \\ 

\sf
\noi
\begin{tabular}{ll}
Bruce E. Turkington  &  Petr Plech\'a\v{c}  \\
Department of Mathematics and Statistics  &  
Department of Mathematics   \\ 
University of Massachusetts  &  University of Tennessee \\
Amherst, MA 01003  & Knoxville, TN 37996-1300 \\  
  &  \\
 \vspace{3mm}   turk@math.umass.edu    &  
  \vspace{3mm}   plechac@math.utk.edu    
\end{tabular}

\end{document}